\documentclass[twocolumn]{aastex701}[linenumbers,twocolumn]
\DeclareUnicodeCharacter{02BC}{\-}
\usepackage{textcomp, gensymb}
\usepackage{booktabs}
\usepackage{soul}
\usepackage{float}
\usepackage{comment}
\usepackage{array}
\usepackage{makecell}
\usepackage{amsmath}
\usepackage{multirow} 
\usepackage{enumitem}

\begin{document}

\title{The First Measurement of Jet Collimation Profiles in an X-ray Binary: the Case of SS\,433}

\correspondingauthor{Lang Cui}
\email{cuilang@xao.ac.cn}

\author[0009-0003-6680-1628]{Xi Yan}
\affiliation{State Key Laboratory of Radio Astronomy and Technology, Xinjiang Astronomical Observatory, CAS, 150 Science 1-Street, Urumqi 830011, China}
\email{yanxi@xao.ac.cn}

\author[0000-0003-0721-5509]{Lang Cui}
\affiliation{State Key Laboratory of Radio Astronomy and Technology, Xinjiang Astronomical Observatory, CAS, 150 Science 1-Street, Urumqi 830011, China}
\affiliation{Xinjiang Key Laboratory of Radio Astrophysics, 150 Science 1-Street, Urumqi, Xinjiang, 830011, China}
\email[]{cuilang@xao.ac.cn}

\author[0000-0002-5195-335X]{Zsolt Paragi}
\affiliation{Joint Institute for VLBI ERIC, Oude Hoogeveensedijk 4, 7991 PD Dwingeloo, The Netherlands}
\email[]{zsolt.paragi@gmail.com}

\author[0000-0003-3079-1889]{S\'andor Frey}
\affiliation{Konkoly Observatory, HUN-REN Research Centre for Astronomy and Earth Sciences, Konkoly Thege Mikl\'os \'ut 15-17, H-1121 Budapest, Hungary}
\affiliation{CSFK, MTA Centre of Excellence, Konkoly Thege Mikl\'os \'ut 15-17, H-1121 Budapest, Hungary}
\affiliation{Department of Astronomy, Institute of Physics and Astronomy, ELTE Eötvös Loránd University, P\'azm\'any P\'eter s\'et\'any 1/A, H-1117 Budapest, Hungary}
\email{frey.sandor@csfk.org}

\begin{abstract}

While significant progress has been achieved in active galactic nuclei (AGN), jet collimation profiles in X-ray binaries (XRBs) have not been directly measured. Here we report the first measurement of jet collimation profiles in an XRB using very long baseline interferometry data from SS\,433. The approaching jet exhibits a well-constrained quasi-parabolic profile in the 1995 and 1998 data, whereas the prominent local oscillations present in the 2000 jet-width measurements preclude a robust characterization of the jet collimation profile. Nevertheless, the 2000 data suggest that the intrinsic jet opening angle decreases gradually from $\sim 20^\circ$ at 8$\times 10^{14}$\,cm to $\sim 3^\circ$ at $9\times 10^{15}$\,cm (deprojected), providing evidence for progressive jet collimation. The width and opening angle of the receding jet are also presented, although their interpretation is limited by the free--free absorption effect. In addition, we measure the frequency-dependent core positions at four bands and derive a core-shift relation. These results establish SS\,433 as the first XRB in which jet collimation, opening-angle evolution, and core-shift relation are constrained. 

\end{abstract}


\section{Introduction}  \label{sec:Introduction}

\begin{deluxetable*}{ccccccccccc}[htb!]
\tablecaption{Summary of Observations and Data of SS\,433 \label{table:SS433_observation_summary}}
\tablehead{\colhead{Date} & \colhead{Code} & \colhead{Array$^a$} & \colhead{$\nu$} & \colhead{$\Theta_{\rm maj} \times \Theta_{\rm min}$, PA} & \colhead{$I_{\rm peak}$} & \colhead{$I_{\rm rms}$} & \colhead{$S_{\rm tot}$} & \colhead{$\Psi_{\rm prec}$} & \colhead{$\theta_{\rm view}$} & \colhead{PA$_{\rm jet}$} \\
& & & (GHz) & (mas $\times$ mas, $\degr$) & \multicolumn{2}{c}{(mJy\,beam$^{-1}$)} & (mJy) & & \multicolumn{2}{c}{($\degr$)} \\
\cmidrule(r){6-7}  \cmidrule(r){10-11} 
(1) & (2) & (3) & (4) & (5) & (6) & (7) & (8) & (9) & (10) & (11) 
}
\startdata
1995/05/06 & BV005 & VLBA+Y1 & 1.7 & $13.3\times7.49, -4.3$ & 42.3 & 0.30 & $286\pm29$ & 0.02 & $\sim 57$ & $95\pm2$ \\
.... & ... & ... & 5 & $3.84\times1.86, -3.8$ & 24.0 & 0.30 & $145\pm15$  & ... & ... & ... \\
\hline
1998/06/16 & BP042d & VLBA+Y1 & 5 & $3.61\times1.47, -4.1$ & 51.2 & 0.40 & $278\pm28$ & 0.02 & $\sim 57$  & $104\pm1$ \\
... & ...  & ... & 8.4 & $2.91\times1.46, -1.9$ & 76.0 & 0.50 & $247\pm25$ & ... & ... & ... \\
... & ... & ... & 15 & $1.65\times1.33, 25.2$ & 77.7 & 0.50 & $193\pm19$ & ... & ... & ... \\
\hline
2000/02/13 & GP025a & VLBA+EVN & 1.7 & $5.68 \times 5.12, -14$ & 30.0 & 0.13 & $250\pm25$ & 0.76 & $\sim 77$ & $81 \pm 1$ \\
2000/02/20 & GP025b & VLBA+EVN & 1.7 & $6.62 \times 4.97, -11$ & 21.2 & 0.16 & $202 \pm 20$ & 0.81 & $\sim 72$ &  $ 80 \pm 1$ \\
2000/05/27 & GP025c & VLBA+EVN & 1.7 & $7.11 \pm 3.68, -11$ & 22.0 & 0.12 & $274 \pm 27$ & 0.40 & $\sim 95$ & $116 \pm 1$ \\
\enddata
\tablecomments{
Columns~(1)--(4): observing date, project code, VLBI array and frequency. 
Column~(5): major axis, minor axis, and position angle of the synthesized elliptical Gaussian beam.
Columns~(6)--(8): peak intensity, rms noise level, and total flux density of the image, respectively. 
Columns~(9) and (10): precession phase and viewing angle of the jet, derived from the well-established kinematic model of SS\,433 \citep[e.g.,][]{Hjellming_1981ApJ...246L.141H,Eikenberry_2001ApJ...561.1027E}, as described in Section~\ref{Results:jet_VA}. 
Column~(11): position angle of the jet, measured from the images shown in \autoref{fig:SS433_images}.
\flushleft $^a$ 
Y1 denotes a single element of VLA.
Eight EVN stations participated in the 2000 observations: Effelsberg, Hartebeesthoek, Jodrell Bank, Medicina, Noto, Onsala, Toru\'n, and Westerbork.
For the 1998 observations, no fringes were detected at the Hancock station at 8.4\,GHz, or at the Hancock, Mauna Kea, and St. Croix stations at 15\,GHz. The 15\,GHz data were used exclusively for determining the core shift, rather than for jet width measurements.
}
\end{deluxetable*}

How relativistic jets in accreting compact objects are collimated is an open question in astrophysics. Although significant progress has been made in active galactic nuclei (AGN) in recent years \citep[e.g.,][]{Asada_2012ApJ...745L..28A,Giovannini_2018NatAs...2..472G,Baczko_2024A&A...692A.205B,Yan_2023,Yan_2025ApJ...991...75Y}, jet collimation profiles in X-ray binaries (XRBs) remain unexplored. A major reason is that steady jets in the hard state are typically too faint and compact to be sufficiently detected, whereas transient jets often appear as a sequence of discrete ejecta rather than continuous flows \citep[e.g.,][]{Fender_2014SSRv..183..323F}. The lack of precise core-shift measurements has also hindered efforts to trace the jet width as a function of distance from the central accretor. 

For AGN, one particularly interesting result is the discovery of well-collimated jets in some high-Eddington systems. For example, the jet in 3C\,273 remains collimated out to $\sim 200$\,pc \citep{Okino_2022ApJ...940...65O}; more remarkably, the jet in CTA\,102 maintains collimation even on scales of $> 1000$\,pc \citep{Ng_2025MNRAS.542..417N}. These findings hint that well-collimated jets may also be observable in Galactic XRBs accreting at high Eddington ratios. In this paper, we report the first measurement of jet collimation profiles in an XRB using very long baseline interferometry (VLBI) data from the well-known hyper-Eddington system SS\,433 \citep{Stephenson_1977ApJS...33..459S}.

SS\,433 is embedded within the giant radio nebula W\,50 \citep{Dubner_1998AJ....116.1842D}. It is particularly notable for the detection of highly red- and blueshifted optical emission lines \citep[for a early review see][]{Margon_1984ARA&A..22..507M}. This peculiar behavior is well explained by a pair of collimated mildly-relativistic jets moving at a speed of $\sim 0.26\,c$ and precessing with a period of $\sim 162$ days, as described by the ``kinematic" model \citep{Fabian_1979MNRAS.187P..13F,Milgrom_1979A&A....76L...3M,Abell_1979Natur.279..701A}.

Subsequent extensive radio observations have independently confirmed the precessing bipolar jets in SS\,433. On arcsecond scales, observations with Very Large Array (VLA) revealed a characteristic corkscrew/zigzag morphology in the twin jets, which closely follows the trajectory predicted by the kinematic model \citep[e.g.,][]{Hjellming_1981Natur.290..100H,Hjellming_1981ApJ...246L.141H,Blundell_2004ApJ...616L.159B,Miller-Jones_2008ApJ...682.1141M}. On milliarcsecond (mas) scales, VLBI observations directly unveiled periodic variations in the jet orientation \citep{Niell_1981ApJ...250..248N,Vermeulen_1987Natur.328..309V,Vermeulen_1993A&A...270..177V,Jeffrey_2016MNRAS.461..312J}. Interestingly, an equatorial outflow oriented nearly perpendicular to the jet axis has also been detected \citep{Paragi_1999A&A...348..910P, Blundell_2001ApJ...562L..79B}, with additional support from optical and near-infrared emission lines \citep[e.g.,][]{Waisberg_2019A&A...623A..47W}.

Jet precession in SS\,433 has again been proven through Doppler-shifted X-ray emission lines from highly ionized heavy elements \citep{Watson_1986MNRAS.222..261W,Kotani_1994PASJ...46L.147K,Marshall_2002ApJ...564..941M}. Both the jet–disk system and the surrounding W\,50 nebula have been widely studied in the X-ray band \citep[e.g.,][]{Brinkmann_1996A&A...312..306B,Safi_1997ApJ...483..868S,Brinkmann_2007A&A...463..611B,Tsuji_2025ApJ...993L..24T,Sunyaev_2026A&A...707A.278S}. At higher energies, the SS\,433/W\,50 complex represents the first XRB detected in the very-high-energy ($E>100$\,GeV) regime \citep{Abeysekara_2018Natur.562...82A}. More recently, ultra-high-energy ($E>100$\,TeV) emission has also been detected by LHAASO, suggesting efficient particle acceleration and a hadronic origin \citep{LHAASO_2025NSRev..12af496L}. As of now, SS\,433 and 4U1630--47 are the only two known XRBs reported to host baryonic jets \citep{Diaz_2013Natur.504..260D}, although the evidence for baryonic jets in the latter remains debated \citep[e.g.,][]{Wang_2016MNRAS.456.1579W}.

SS\,433 is also notable as a hyper-Eddington accretion system \citep[for a review see][]{Fabrika_2004ASPRv..12....1F}. Its central compact object is most likely a black hole with a mass of $\gtrsim 8\,M_{\odot}$, accompanied by an A-type supergiant donor star \citep{Cherepashchuk_2025PhyU...68.1042C}. At a distance of $5.5 \pm 0.2$\,kpc \citep{Blundell_2004ApJ...616L.159B}, we obtain $1\,\mathrm{mas} \leftrightarrow 5.5\,\mathrm{AU} \leftrightarrow 8.2\times 10^{13}\,\mathrm{cm}$. 
In this work, we study the jet collimation behavior of SS\,433 using archival VLBI data. Section~\ref{sec:Observation_and_image_analysis} describes the observations and data analysis. Section~\ref{sec:Results} presents the results we obtained, which are discussed in Section~\ref{sec:Discussion}. Our main findings are summarized in Section~\ref{sec:Summary}.

\begin{figure*}[htbp!]
\centering
    \hspace{-0.4cm}
    \includegraphics[width=0.325\linewidth]{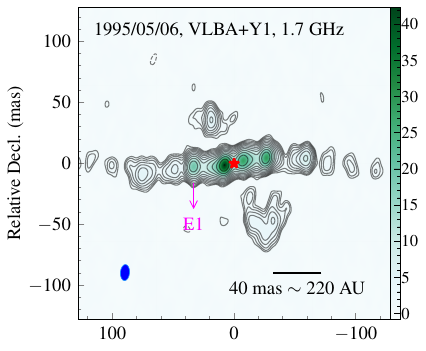}
    \includegraphics[width=0.3\linewidth]{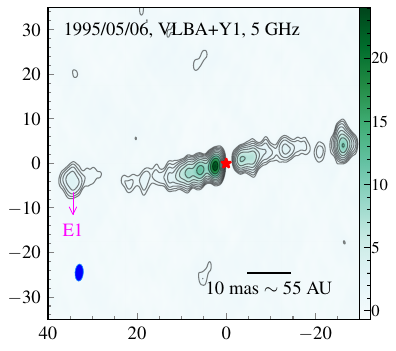}
    \hspace{5cm}
    \includegraphics[width=0.325\linewidth]{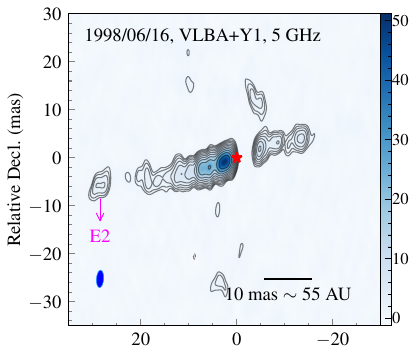}
    \includegraphics[width=0.31\linewidth]{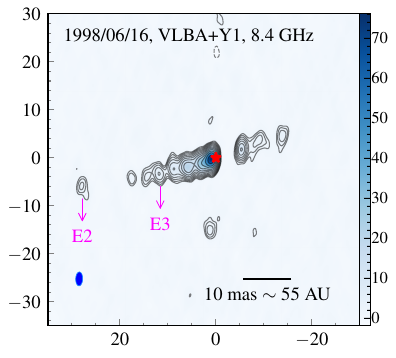}
    \includegraphics[width=0.31\linewidth]{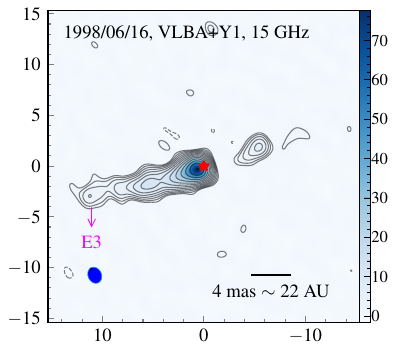}
    \includegraphics[width=0.50\linewidth]{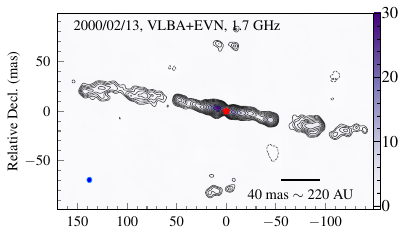}
    \includegraphics[width=0.48\linewidth]{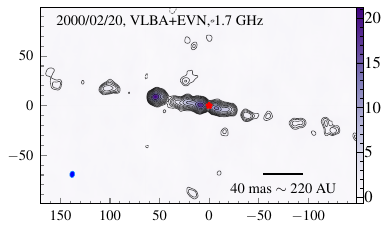}
    \includegraphics[width=0.52\linewidth]{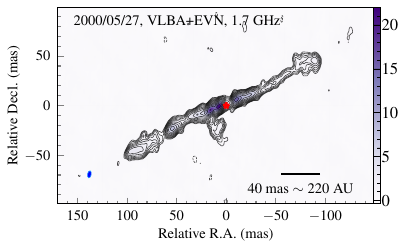}
\caption{
VLBI images of SS\,433 obtained from different projects (see \autoref{table:SS433_observation_summary}). The color bar indicates the intensity in mJy\,beam$^{-1}$, and the contour levels are $\pm 3 I_{\rm rms} \times \sqrt{2}^{\,n}$, with $n = 1, 2, 3, \ldots$. The synthesized beam is shown by the blue ellipse in the bottom-left corner of each panel. The optically thin E1, E2, and E3 components are used to determine the core shifts for different frequency pairs (see Section~\ref{Results:core_shift}). The putative location of the central binary system has been shifted to the image center based on the measured core shifts (see \autoref{table:core_shift}) and is marked by a red star. A gap is apparent in these images, which is attributed to free–free absorption (FFA) affecting the western jet emission \citep{Paragi_1999A&A...348..910P}.
}
\label{fig:SS433_images}
\end{figure*}

\section{Observations and Image Analysis} \label{sec:Observation_and_image_analysis}
\subsection{Observations and Images} \label{subsec:Observations} 

We analyzed archival VLBI data of SS\,433 obtained with Very Long Baseline Array (VLBA) together with either a single antenna of VLA, or stations from European VLBI Network (EVN) (see \autoref{table:SS433_observation_summary}).\footnote{\url{https://data.nrao.edu/}.} The data were calibrated in the Astronomical Image Processing System \citep[{\tt AIPS};][]{Greisen2003} following standard procedures described in the {\tt AIPS} cookbook (see Chapter~9).\footnote{\url{https://www.aips.nrao.edu/cook.html}.} Imaging with iterative phase/amplitude self-calibration was performed in {\tt DIFMAP} \citep{Shepherd_Difmap_1997ASPC..125...77S}. The resulting CLEAN images of SS\,433 are shown in \autoref{fig:SS433_images}, and are consistent with those presented in \citet{Paragi_1999A&A...348..910P, Paragi_2002evn..conf..263P, Paragi_2003nvm..conf..261P} (for detailed discussion on the images, see these references).

\subsection{Image Analysis}  \label{subsec:Image_analysis}

To measure the deconvolved jet width of SS\,433, we followed the approach described in \citet{Park_2021ApJ...909...76P}, which is commonly used in jet collimation studies of AGN. First, the location of the binary system was shifted to the image center based on the measured core shift (see Section~\ref{Results:core_shift}). The image was then restored with a circular beam whose size was set equal to the major axis of the synthesized beam ($\Theta_{\rm maj}$; see \autoref{table:SS433_observation_summary}) and rotated such that the jet axis was aligned along the east--west direction. 

At each projected distance from the central binary ($r_{\rm proj}$), we extracted the transverse intensity profile of the jet using a pixel-based approach. We note that the north--south equatorial structures (see \autoref{fig:SS433_images}) were excluded prior to this extraction. The transverse intensity profiles were then fitted with a Gaussian function to derive the full width at half maximum (FWHM), $\Theta_{\rm fit}$. A measurement was retained only if (1) the fitted Gaussian amplitude exceeds $20\,I_{\rm rms}$ ($I_{\rm rms}$ is the image rms noise; see \autoref{table:SS433_observation_summary}), and (2) $\Theta_{\rm fit} > \Theta_{\rm maj}$. The deconvolved jet width was then derived as $W = ({\Theta_{\rm fit}^2 - \Theta_{\rm maj}^2})^{1/2}$.

Due to the resolution limitations, we only retained measurements at distances of $\gtrsim\Theta_{\rm maj}$ from the central core. Then, the deconvolved jet widths were binned as a function of distance using a bin size of $\Theta_{\rm maj}/5$. For each bin, the mean value was adopted as the representative jet width, assuming an uncertainty of $\Theta_{\rm maj}/10$.\footnote{We measured the jet width only for Gaussian components with amplitudes exceeding $20\,I_{\rm rms}$. This corresponds to a nominal uncertainty of approximately $d/(\mathrm{S/N})$ \citep[e.g.,][]{lee_2008AJ....136..159L}, where $d$ is the component size and S/N is the signal-to-noise ratio, which is $\geq 20$ in our analysis. Therefore, the adopted uncertainty of $\Theta_{\rm maj}/10$ is more conservative than this estimate. }

Using the derived jet width $W$, the projected distance from the binary system $r_{\rm proj}$, and the corresponding jet viewing angle $\theta_{\rm view}$ (see Section~\ref{Results:jet_VA}), we calculated the intrinsic jet opening angle:
\begin{equation} \label{eq:eq1}
    \phi_{\rm open} = 2\,\arctan\left(\frac{W\,\sin\theta_{\rm view}}{2\,r_{\rm proj}}\right).
\end{equation}

\begin{deluxetable}{ccccc}[t!]
\tablecaption{Core Positions of the Approaching Jet Relative to the Central Binary in R.A. and Decl. (mas) \label{table:core_shift}}
\tablehead{ & \colhead{1.7\,GHz} & \colhead{5\,GHz} & \colhead{8.4\,GHz} & \colhead{15\,GHz}}
\startdata
$\Delta{\rm R.A.} $ & $7.9\pm1.7$  & $2.5\pm0.8$  & $1.0\pm0.9$ & $0.7\pm0.7$  \\
$\Delta{\rm Decl.}$ & $-1.6\pm1.7$ & $-0.7\pm0.8$ & $-0.4\pm0.9$ & $-0.3\pm0.7$\\
\enddata
\end{deluxetable}

\section{Results} \label{sec:Results}
This study focuses on the jet collimation profile of SS\,433, i.e., the deconvolved jet width as a function of deprojected distance from the central binary system. Two effects must be taken into account. First, the apparent core position is frequency dependent (i.e., core shift) due to the synchrotron self-absorption effect. Second, the precession of the SS\,433 jet causes its viewing angle to vary with time. Therefore, prior to constructing the jet collimation profile, it is necessary to accurately determine the location of the binary system and the jet viewing angle for each observing epoch.

\subsection{Core Shift on the Eastern Side} \label{Results:core_shift}
Following the method adopted in \citet{Paragi_1999A&A...348..910P}, we measured the core shift in the approaching jet using optically thin jet components. As shown in \autoref{fig:SS433_images}, the E1 component in the 1995 images was used to derive the relative core separation between 1.7 and 5\,GHz. For the 1998 observations at higher frequencies, the E2 and E3 components were used to determine the core shifts for the 5--8.4\,GHz and 8.4--15\,GHz frequency pairs, respectively. Using the well-established kinematic model of SS\,433, \citet{Paragi_1999A&A...348..910P} found that the most plausible location of the kinematic center (i.e., the binary system) is approximately 2.5\,mas west and 0.7\,mas north of the eastern core at 5\,GHz. Adopting this constraint, we determined the core positions relative to the central binary at each frequency, as summarized in \autoref{table:core_shift}.\footnote{Independently, we also measured the core shift using a two-dimensional cross-correlation analysis implemented in the VIMAP package \citep{Kim_2014JKAS...47..195K}. The results are consistent with those listed in \autoref{table:core_shift}.}

As seen, the core shift is mainly along the R.A. direction. Notably, we find no significant core offset at 8.4 and 15\,GHz. This may indicate that the core positions at these frequencies are located close to the binary, which cannot be reliably resolved with our observations. In Section~\ref{Discussion:core_shift}, we will use these measurements to constrain the core-shift relation.

\subsection{Jet Viewing Angle} \label{Results:jet_VA}
We derived the jet viewing angle using the well-established kinematic model of SS\,433, which is characterized by the intrinsic speed of $\beta \approx 0.26$, the inclination of the precession axis, $i = 78\fdg05 \pm 0\fdg05$, the half-opening angle of the precession cone, $\psi = 20\fdg92 \pm 0\fdg08$, and the precession period, $P_{\rm prec} = 162.375 \pm 0.011$ days \citep{Eikenberry_2001ApJ...561.1027E}.

The precession phase is defined as $\Psi_{\rm prec} = (T - T_{0})/P_{\rm prec}$, where $T$ is the Julian Date (JD) of the observation and $T_{0} = \mathrm{JD}\,2443507.47$ corresponds to zero precessional phase, at which the eastern approaching jet is closest to the line of sight \citep{Fabrika_2004ASPRv..12....1F}. Based on these parameters, the jet viewing angle can be calculated using \citep[e.g.,][]{Hjellming_1981ApJ...246L.141H}:
\begin{equation} \label{eq:eq2}
    \cos\theta_{\rm view} = \sin i \, \sin \psi \, \cos\bigl(2\pi \, \Psi_{\rm prec}\bigr) + \cos i \, \cos \psi .
\end{equation}
The derived precession phase and jet viewing angle for each epoch are listed in \autoref{table:SS433_observation_summary}.

\subsection{Jet Collimation Profiles} \label{Results:jet_collimation_profiles}
\subsubsection{1995 and 1998 Datasets}
In addition to the images presented in \autoref{fig:SS433_images}, we made a series of new images for the 1995 and 1998 datasets observed at 1.7, 5, and 8.4\,GHz using different levels of Gaussian $uv$-tapering to recover more diffuse or partially resolved downstream emission. We measured the jet width and opening angle from both the untapered and tapered images. These independently reconstructed images, obtained with different intrinsic resolutions, allow us to cross-check the measurements at individual frequencies and across different frequencies, thereby providing a more reliable characterization of the evolution of the jet width and opening angle.

\begin{figure*}[htbp!]
    \centering
    \includegraphics[width=0.71\linewidth]{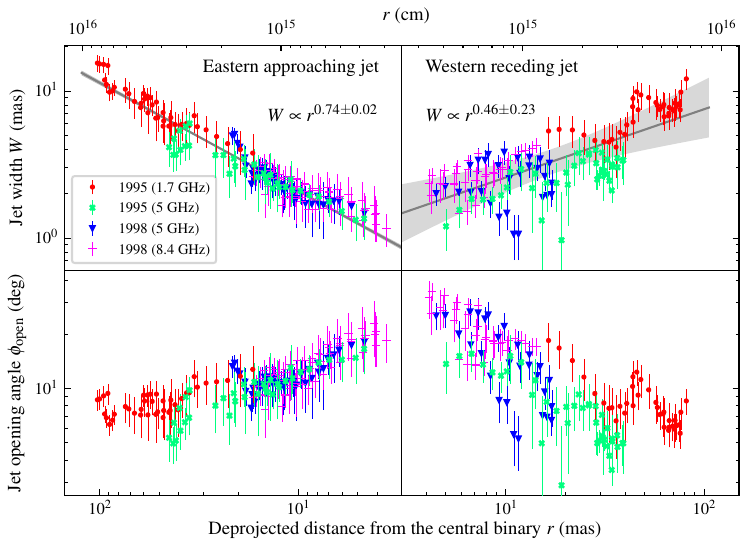}
    \caption{
    Deconvolved jet width (upper panel) and intrinsic jet opening angle (lower panel) as a function of deprojected distance from the central binary on the approaching and receding sides of SS\,433, derived from 1995 and 1998 datasets. The lower horizontal axis shows the deprojected distance in mas, while the upper horizontal axis shows the corresponding distance in cm, adopting a source distance of 5.5\,kpc \citep{Blundell_2004ApJ...616L.159B}. The jet viewing angle in both epochs is derived to be $\sim 57^\circ$ (see Section~\ref{Results:jet_VA}). The gray solid line shows the best power-law fit, with the shaded region indicating the $1\sigma$ confidence interval. 
    }
\label{fig:SS433_JC_1995and1998}
\end{figure*}

In \autoref{fig:SS433_JC_1995and1998}, we present the deconvolved jet width and intrinsic opening angle of the approaching and receding jets in SS\,433 as a function of deprojected distance from the central binary. Despite the three-year interval between the 1995 and 1998 observations, the measurements from the two epochs are consistent with each other. As shown in \autoref{fig:SS433_images}, the overall jet morphologies are also remarkably similar. 

To characterize the jet collimation profile, we fitted the jet width with a power-law relation of the form $W \propto r^a$, where $a = 0$, $0.5$, and $1$ correspond to cylindrical, parabolic, and conical geometries, respectively. The approaching jet is well described by a quasi-parabolic profile ($W \propto r^{0.74 \pm 0.02}$) over deprojected distances of $\sim 4 - 100$\,mas, corresponding to $\sim 3 \times 10^{14} - 8 \times 10^{15}$\,cm. In comparison, the receding jet follows a parabolic profile ($W \propto r^{0.46 \pm 0.23}$) over a comparable physical scale. However, the width measurements of the receding jet exhibit substantially larger scatter than those of the approaching jet. This is primarily because the receding jet is affected by FFA \citep{Paragi_1999A&A...348..910P}, as can be seen from \autoref{fig:SS433_images}. In view of this, we present the results for the receding jet but do not discuss them in detail.

Our results also show that the intrinsic opening angle of both jets decreases from $\sim 20^\circ$ -- $30^\circ$ at a deprojected distance of $\sim4$\,mas to a few degrees at $\sim 100$\,mas, indicative of progressive collimation of the bipolar outflow.

\subsubsection{2000 Datasets}
\begin{figure*}[htbp!!]
    \centering
    \includegraphics[width=0.69\linewidth]{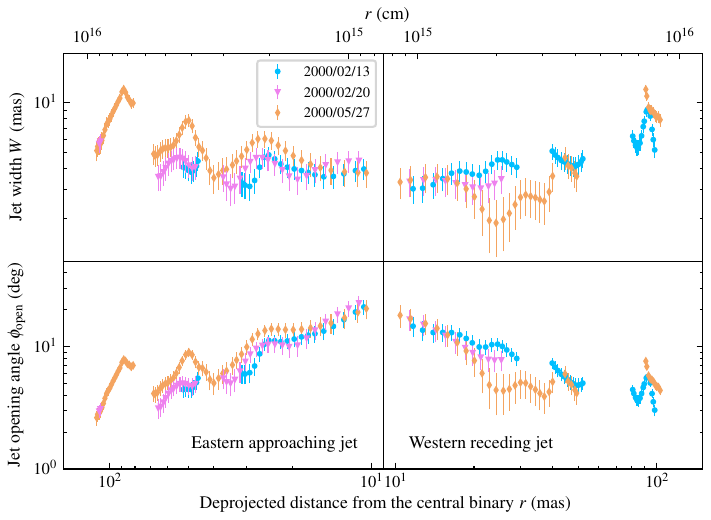}
    \caption{Deconvolved jet width (upper panel) and intrinsic jet opening angle (lower panel) as a function of deprojected distance from the central binary on the approaching and receding sides of SS\,433, derived from the three epochs in 2000. The jet viewing angle at each epoch is listed in \autoref{table:SS433_observation_summary}. 
    }
\label{fig:SS433_JC_combined_2000}
\end{figure*}

Both the angular resolution and sensitivity of the 1.7\,GHz observations in 2000 are improved compared to the 1995 observations at the same frequency, thanks to the participation of EVN stations (see \autoref{table:SS433_observation_summary}). The measured jet width and opening angle as a function of deprojected distance from the binary, obtained by combining the three epochs in 2000, are shown in \autoref{fig:SS433_JC_combined_2000}. 

Several features are evident. First, the approaching jet widths are almost consistent across the three epochs, whereas the receding jet shows local discrepancies. Second, the width profiles of both jets show a clear transition from a flatter to a steeper profile. Finally, prominent oscillatory features (or jet ``humps'') are present in the width profiles. These features prevent us from robustly characterizing the jet collimation profile with either a single power-law or a broken power-law fit. Nevertheless, both jets show reliable evidence of collimation, as the intrinsic opening angle of the approaching (receding) jet decreases from $20^\circ \pm 4^\circ$ ($18^\circ \pm 4^\circ$) at $\sim10$\,mas to $3^\circ \pm 1^\circ$ ($4^\circ \pm 1^\circ$) at $\sim112$\,mas. 

\begin{figure}[htbp!]
\centering
    \centering
    \includegraphics[width=0.95\linewidth]{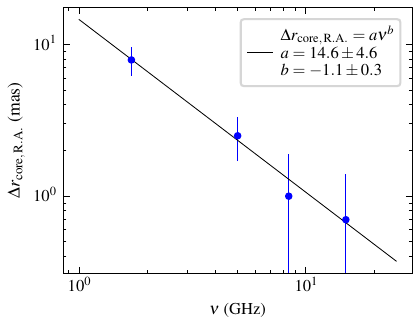}
    \caption{Approaching jet core position relative to the central binary in R.A. as a function of frequency. The black solid line shows the best-fit power law.
    }
\label{fig:SS433_core_shift}
\end{figure}

\section{Discussion} \label{sec:Discussion}
\subsection{Frequency-dependent Core Shift} \label{Discussion:core_shift}
In comparison with AGN, studies of frequency-dependent core shifts in Galactic XRBs remain limited. SS\,433 was the first source in which the core shift was measured \citep[across three frequency bands;][]{Paragi_1999A&A...348..910P}. More recently, core shifts between two frequency bands have also been reported for V404 Cygni, MAXI J1820+070, and Swift J1727.8--1613 \citep{Prabu_2023MNRAS.525.4426P,Wood_2025ApJ...984L..53W}. In this work, we extend these studies by measuring the core shift of the approaching jet in SS\,433 at 1.7, 5, 8.4, and 15\,GHz.

Using the results listed in \autoref{table:core_shift}, we explored the dependence of the core position on observing frequency in SS\,433. As shown in \autoref{fig:SS433_core_shift}, a power-law fit yields $\Delta r_{\rm core,\,R.A.} \propto \nu^{-1.1 \pm 0.3}$. This derived index is consistent with the theoretical expectation of $-1$ for synchrotron self-absorbed jets \citep[e.g.,][]{Konigl_1981ApJ...243..700K}, and is comparable to values reported for nearby AGN \citep[e.g.,][]{Hada_2011Natur.477..185H, Haga_2015ApJ80715H, Park_2021ApJ...909...76P, Boccardi_2021A&A...647A..67B, Yan_2025ApJ...983..169Y}. SS\,433 thus represents the first Galactic XRB in which a core-shift relation has been constrained.

\begin{figure*}[htbp!]
    \centering
    \includegraphics[width=0.9\linewidth]{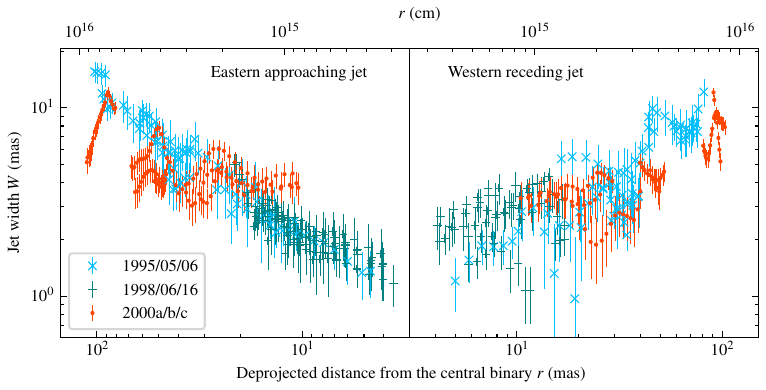}
    \caption{
    Collimation profiles of both the approaching and receding jets in SS\,433, obtained by combining measurements from 1995, 1998, and 2000 (see Figures~\ref{fig:SS433_JC_1995and1998} and \ref{fig:SS433_JC_combined_2000}).
    }
\label{fig:SS433_JC_combined_199519982000}
\end{figure*}

\subsection{Diverse Jet Collimation Behaviors?} 
\label{Discussion:jet_collimation}
Compared to the well-constrained quasi-parabolic profile of the approaching jet in 1995 and 1998 (see \autoref{fig:SS433_JC_1995and1998}), the jet width measurements obtained in 2000 exhibit several prominent humps (see \autoref{fig:SS433_JC_combined_2000}). Similar oscillatory features have also been detected in AGN jet width profiles; however, their physical origin remains unclear \citep{Ng_2025MNRAS.542..417N}. Future multi-frequency, high-quality VLBI observations may help to unveil the origin of these jet-width humps in SS\,433.

Irrespective of these jet-width humps, \autoref{fig:SS433_JC_combined_199519982000} presents the combined jet-width measurements from 1995 to 2000. Despite being obtained with different arrays (see \autoref{table:SS433_observation_summary}), these measurements remain overall consistent, although local, non-systematic differences are also present. This suggests that resolution effects from the different arrays are likely not significant.

A possible explanation for the observed width differences between the 1995/1998 and 2000 data is intrinsically different collimation behaviors in SS\,433, which may be supported by two aspects. First, similar differences in the width evolution between the 1995/1998 and 2000 datasets are also hinted in model-fitted FWHMs derived from {\tt DIFMAP}. In addition, as shown in \autoref{appendix:1999_jet_collimation_profile}, the 1999 VLBA+VLA observation at 1.4\,GHz shows a roughly cylindrical jet on 40--270\,mas scales, in contrast to the geometries derived from the 1995/1998 and 2000 data. In this context, we may reveal diverse jet collimation behaviors in SS\,433. In particular, we note that such diversity in jet geometry is unlikely in a single AGN, as AGN evolutionary timescales (millions of years) are far longer than those of XRBs (weeks to months).

Beyond SS\,433, the number of currently available XRBs suitable for jet collimation studies is very limited. Potential candidates include persistent systems such as GRS\,1915+105\footnote{Notably, it has recently been reported that GRS\,1915+105 has faded into a very low-luminosity or even quiescent state \citep{Marino_2026ATel17869....1M,Motta_2026ATel17865....1M,gandhi2026intrinsicdeclineaccretionactivity}, which would make such observations challenging.} \citep[e.g.,][]{Dhawan_2000ApJ...543..373D,Miller-Jones_2007MNRAS.375.1087M,Yan_2026}, Cygnus\,X-1 \citep[e.g.,][]{Prabu_2026NatAs.tmp...83P}, and Cygnus\,X-3 \citep{Mioduszewski_2001ApJ...553..766M,Miller-Jones_2004ApJ...600..368M}, as well as the transient source J1727.8$-$1613, which exhibits the largest resolved continuous jet ever observed in an XRB \citep{Wood_2024ApJ...971L...9W}. However, with the advent of next-generation interferometric arrays, it may become possible to investigate jet collimation in XRBs in much greater detail. More importantly, comparisons with AGN may provide deeper insight into jet collimation mechanisms in accreting black hole systems across vastly different mass scales.

\subsection{Jet Opening Angle} \label{Discussion:Jet_opening_angle}
The opening angles of SS\,433’s X-ray, optical, and radio jets have been measured in previous studies \citep[e.g.,][]{Fabrika_2004ASPRv..12....1F}. The X-ray jet, observed on scales of $10^{10}$--$10^{13}$\,cm, is tightly collimated to $\sim 1^\circ$--$6^\circ$ \citep{Marshall_2002ApJ...564..941M, Marshall_2013ApJ...775...75M, Fogantini_2023A&A...669A.149F}, likely constrained by the disk wind and hot gas cocoons extending up to $\lesssim 10^{13}$\,cm (see Fig.~13 of \citealt{Fabrika_2004ASPRv..12....1F} for an illustration of the SS\,433 binary system and jets).

Beyond $\sim 10^{13}$\,cm, where the disk wind and cocoon cease to dominate, the jet is expected to break out of the strong confinement that shapes the inner X-ray jet, allowing it to expand transversely into the surrounding medium. Consistent with this picture, our VLBI observations reveal a substantially larger opening angle of $\sim 20^\circ$ at (3–8)$\times 10^{14}$\,cm. However, the jet does not subsequently evolve into a freely expanding outflow. Instead, the opening angle gradually decreases with distance, reaching only $\sim 3^\circ$ at $9\times 10^{15}$\,cm, suggesting that the jet undergoes progressive collimation on these scales. This behavior naturally connects to earlier findings that SS\,433 remains confined until transitioning to free expansion only at $\sim 6.5\times 10^{16}$\,cm \citep[][see also Section~\ref{Discussion:Jet_confinement}]{Hjellming_1988ApJ...328..600H}, highlighting the complex interplay between the jet and its surrounding environment over several orders of magnitude in scale.

For the optical jet, the opening angle was measured to be $\sim 1^\circ$--$5^\circ$ \citep{Begelman_1980ApJ...238..722B, Borisov_1987SvAL...13..200B}, which is consistent with our measurements, given that the optical jet extends over $\gtrsim 10^{14}$\,cm, similar to the radio jet. Finally, we note that studies of the jet opening angle in SS\,433 on VLBI scales remain sparse. For example, early VLBI observations suggested an upper limit of $15^\circ$, based on visibility amplitude fitting \citep{Niell_1981ApJ...250..248N}. In this work, our results not only provide direct measurements of the intrinsic jet opening angle from the images, but also reveal how it evolves with increasing distance from the central binary.

\subsection{Jet Confinement} \label{Discussion:Jet_confinement}
There is compelling evidence for jet confinement in SS\,433. \citet{Hjellming_1988ApJ...328..600H} analyzed the flux density decay of a single impulsive ejection and found a transition from $S_{\nu} \propto t^{1.3}$ to $S_{\nu} \propto t^{2.5}$. Based on the models, they explained this behavior as indicative of the jet undergoing a transition from slowed (confined) to free (unconfined) expansion at a distance of $\sim 6.5 \times 10^{16}$\,cm. On the other hand, assuming the jet can expand transversely at the speed of light, \citet{Miller-Jones_2006MNRAS.367.1432M} estimated a jet opening angle of at least $74^\circ$ for SS\,433. This estimate sharply contradicts observations, leading to the conclusion that the jets in SS\,433 are certainly confined. In this study, we present the most direct evidence to date for gradual jet confinement in SS\,433, based on the jet collimation profiles and the radial evolution of the intrinsic opening angle (e.g., Section~\ref{Results:jet_collimation_profiles}).

The jet confinement mechanisms in XRBs have been discussed in detail by \citet{Miller-Jones_2006MNRAS.367.1432M}. First, the magnetic confinement is likely to play an important role. Polarimetric observations with the Atacama Large Millimeter/submillimeter Array at 230\,GHz have shown that the magnetic field vectors are oriented perpendicular to the jet ridge line at distances of $\lesssim 350$\,mas (corresponding to $\sim 3 \times 10^{16}$\,cm); beyond this region, however, the magnetic field becomes aligned parallel to the ridge line in both jets (\citealt{Blundell_2018ApJ...867L..25B}; see also \citealt{Stirling_2004MNRAS.354.1239S} and \citealt{Miller-Jones_2008ApJ...682.1141M} for discussions of the magnetic field orientation in the jets of SS\,433). If the perpendicular configuration corresponds to a toroidal magnetic field, the jet could be confined by magnetic hoop stress. Notably, the transition zone in magnetic field direction broadly coincides with the location where the radio flux density decay changes (see above).

On the other hand, SS\,433 is quite unique in exhibiting ultra-high-energy emission with a hadronic origin \citep[e.g.,][]{LHAASO_2025NSRev..12af496L}, as well as baryonic jets (see Section~\ref{sec:Introduction}). As discussed by \citet{Miller-Jones_2006MNRAS.367.1432M}, if cold protons dominate the jet inertia, the transverse expansion would be slower than that of a plasma composed solely of electron–positron pairs. In this context, the presence of cold protons provides a plausible explanation for jet confinement in SS\,433, which may also offer valuable insights into jet collimation in AGN.

\subsection{Limitations, Caveats, and Future Prospects}
Although abundant VLA data are available, we do not include them in our analysis of the jet collimation profile. This is due to jet precession. Adopting a mean jet proper motion of $8.3\,\mathrm{mas}\,\mathrm{d}^{-1}$ \citep{Jeffrey_2016MNRAS.461..312J}, the jet completes one full rotation over a projected distance of $\sim 1350$\,mas from the core. As a result, the spiraling jets at distances of several arcseconds encompass more than one precession cycle, producing the observed corkscrew morphology \citep[e.g.,][]{Blundell_2004ApJ...616L.159B}. This effect precludes accurate measurements of the jet width on arcsecond scales. 

On VLBI scales, we measured the jet collimation profile out to $\sim 112$\,mas (at 1.7\,GHz), corresponding to an age of 13.5 days. Given the half-opening angle of the precession cone ($\psi = 20\fdg92$) and the precession period ($P_{\rm prec} = 162.375$\,days) \citep{Eikenberry_2001ApJ...561.1027E}, this implies that the outermost jet has rotated by an angle of $\sim 3\fdg5$. This rotation introduces an apparent width ``increase" of $2 \times 112 \times \mathrm{tan}(3\fdg5/2) \approx 7$\,mas, which is nearly a factor of two smaller than the major-axis size of the synthesized beam for the 1.7\,GHz image in 1995, and comparable to that of the 2000 images (see \autoref{table:SS433_observation_summary}). We note that, prior to measuring the jet width, the images were restored with a circular beam whose size was set equal to the major axis of the synthesized beam (see Section~\ref{subsec:Image_analysis}). Consequently, for the 2000 data, jet rotation may introduce a slight broadening in jet width. However, since this effect is comparable to the synthesized beam, it is unlikely to significantly bias the inferred width profile.
At 5 and 8.4\,GHz, the jet extent is smaller than $20$\,mas; a similar estimate yields an additional apparent broadening of only $\lesssim 0.2$\,mas, which is negligible compared to the synthesized beam size. Therefore, the width measurements at these frequencies are almost unaffected by jet rotation.

We have also carefully searched the NRAO data archive for additional VLBI observations of SS\,433, but find that the data presented in this work are likely the only ones suitable for jet collimation studies. There are several reasons for this: (1) non-detection of the target in some epochs, (2) lack of simultaneous multi-frequency coverage, and (3) in certain cases, the jet appears as multiple isolated blobs during strong radio flares \citep[e.g.,][]{Paragi2012SS433MA,Jeffrey_2016MNRAS.461..312J}, rather than a relatively continuous flow. Future investigations of jet collimation in SS\,433 at higher frequencies (e.g., $\gtrsim 15$\,GHz) will be essential.

\section{Summary} \label{sec:Summary}
Despite significant advances in jet collimation studies of AGN, similar investigations in XRBs has not yet been reported. In this work, we study the collimation behavior of the baryonic jets in SS\,433, a hyper-Eddington system. Our findings are summarized as follows. 

\begin{enumerate}
    \item[-] \textit{Jet collimation profile.} For the approaching jet, we find a well-constrained quasi-parabolic profile using the data from 1995 and 1998. In contrast, the jet width measured from the 2000 data exhibits prominent local oscillations, which preclude a reliable determination of the width profile. We also present the width profiles of the receding jet, but do not interpret these results further, as the emission is affected by free–free absorption. 
    
    \item[-] \textit{Jet opening angle.} The high-resolution VLBI images also enable the first study of the radial evolution of intrinsic jet opening angle in an XRB. Our results show that the opening angle of SS\,433 decreases with increasing distance from the central binary, from $\sim 20^\circ$ at (3--8)$\times 10^{14}$\,cm to $\sim 3^\circ$ at $9\times 10^{15}$\,cm (deprojected). Together with the jet collimation profiles, these results provide strong evidence for progressive jet collimation in SS\,433. We briefly discuss potential jet confinement mechanisms, including contributions from the magnetic field configuration and the presence of cold protons in the jets.

    \item[-] \textit{Core shift.} We extend previous studies of the frequency-dependent core shift in SS\,433 using multi-frequency data at 1.7, 5, 8.4, and 15\,GHz, and derive a relation of $\Delta r_{\rm core,\,R.A.} \propto \nu^{-1.1 \pm 0.3}$. This index is consistent with both theoretical expectations and observations of nearby AGN. SS\,433 thus represents the first Galactic XRB in which a core-shift relation has been constrained.
    
\end{enumerate}

\begin{acknowledgments}
We sincerely thank the referee for the constructive and insightful comments, which greatly improved the quality and clarity of the manuscript. This work is supported by the Tianshan Talent Training Program (grant No. 2023TSYCCX0099) and the National Key R\&D Program of China (grant Nos. 2024YFA1611500 and 2022SKA0120102). 
X.Y. is supported by the China Postdoctoral Science Foundation under grant Nos. 2025M773200 and 2026T190850, and by the Xinjiang Tianchi Talent Program. 
This work was partly supported by the Urumqi Nanshan Astronomy and Deep Space Exploration Observation and Research Station of Xinjiang (XJYWZ2303) and the Central Guidance for Local Science and Technology Development Fund (grant No. ZYYD2026JD01). 

Very Long Baseline Array is operated by the National Radio Astronomy Observatory. The National Radio Astronomy Observatory and Green Bank Observatory are facilities of the U.S. National Science Foundation operated under cooperative agreement by Associated Universities, Inc. 
The European VLBI Network is a joint facility of independent European, African, Asian, and North American radio astronomy institutes. Scientific results from data presented in this publication are derived from the EVN project GP025.

\end{acknowledgments}

\appendix
\section{Jet Collimation Profile from the 1999 VLBA+VLA Observation} \label{appendix:1999_jet_collimation_profile}

\begin{figure*}[htbp!]
    \centering
    \includegraphics[width=0.38\linewidth]{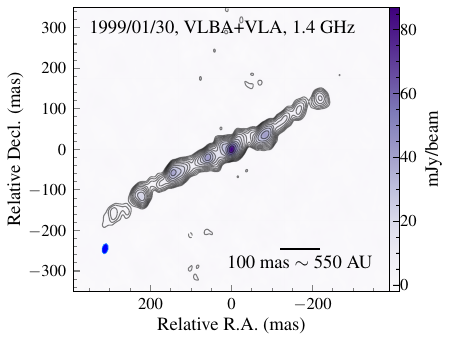}
    \hspace{1cm}
    \includegraphics[width=0.4\linewidth]{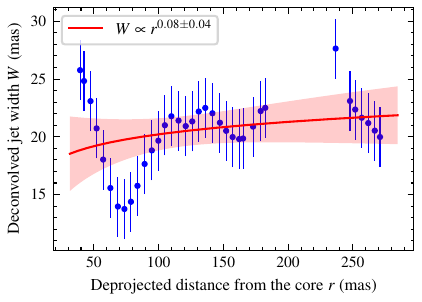}
    \caption{
    \textit{Left:} CLEAN image of SS\,433 observed with VLBA + VLA at 1.4\,GHz. The image was restored using an elliptical Gaussian beam of $26.0 \times 15.2$\,mas, oriented at a position angle of $-12\degr$. Contours start at 1.5\,mJy\,beam$^{-1}$ and increase by factors of $\sqrt{2}$. 
    The jet viewing angle at this epoch is $\sim 97^\circ$. 
    \textit{Right:} deconvolved jet width of the eastern jet as a function of deprojected distance from the core, derived from the image on the left. The red solid line shows the power-law fit, with the shaded region indicating the $3\sigma$ confidence interval. This fit suggests a cylindrical profile.
    }
\label{fig:SS433_JC_1999}
\end{figure*}

SS\,433 was observed with VLBA in combination with all 27 VLA antennas at 1.4\,GHz for $\sim 12.5$ hours on January 30, 1999 (project code: BP050). At this epoch, the precession phase was 0.43, corresponding to a jet inclination of $\sim 97\degr$. The CLEAN image is shown in the left panel of \autoref{fig:SS433_JC_1999}.

Since an emission gap (as seen in the 1.7--15\,GHz images of \autoref{fig:SS433_images}) is not observed/resolved at 1.4\,GHz, we cannot reliably determine the location of the central binary. Consequently, we analyzed the width evolution of the eastern jet as a function of distance from the core, identified as the brightest region in the image, as shown in the right panel of \autoref{fig:SS433_JC_1999}. Despite the presence of local oscillations, the jet appears roughly cylindrical on scales from 40\,mas to 270\,mas. In particular, the measured jet width remains nearly constant at distances $\gtrsim 100$\,mas.

\newpage
\bibliography{SS433}{}
\bibliographystyle{aasjournalv7}

\end{document}